\documentclass[prl,showpacs,showkeys,amsmath,amssymb,twocolumn]{revtex4}

\usepackage{dcolumn}
\usepackage{bm}
\input xy
\usepackage[all]{xy}
\xyoption{2cell} \UseAllTwocells

\begin{document}



\title{Non interacting electron gas model of quantum Hall effect}
\author{Kang Li$^1$}
\email{kangli@hznu.edu.cn}
\author{Shuming Long$^2$}
\email{longshuming@163.com}
\author{Jianhua Wang$^2$}
\email{jianhuawang59@yahoo.com.cn}
\author{Yi Yuan$^2$}
\affiliation{1,Hangzhou Normal University, Hangzhou, 310036, Zhejiang, China}
\affiliation{2,Shaanxi University  of Technology, Hanzhong, 723001,
Shaanxi, China}

\date{\today}

\begin{abstract}
 On the basis of our previous studies on energy levels and wave functions of single electrons in a strong magnetic field, the energy levels and wave functions of non-interacting electron gas system, electron gas Hall surface density and Hall resistance of electron gas system are calculated. Then, a comparison is made between non-interaction electron gas model and Laughlin's interaction two dimensional electron gas system. It is found that the former can more quickly and unified the explain the integer and the fractional quantum Hall effects without the help of concepts proposed by Laughlin, such as fractionally charged quasi-particles and quasi-holes which obey fractional statistics.   After explicit calculation, it is also discovered that quantum Hall effect has the same physical nature as superconducting state.
\end{abstract}

\maketitle

\section{Introduction}

The Hall effect has attracted increasing attention and more and more studies have been done on it in recent years\cite{WLL}-\cite{BMC}. In fact, Hall effect was discovered by  Edwin Hall in 1879. 100 years later, the quantum Hall effect was found\cite{klizing1}-\cite{klizing3}, which is a brilliant achievement in quantum physics. After 1980's, Daniel Chee Tsui, Horst L. Stormer and Robert B.Laughlin made great discoveries on the fractional quantum Hall effect\cite{tsui1}-\cite{laughlin1} and were rewarded the 1998 Nobel Prize in physics.

Until now, a good way to interpret the quantum Hall effect is to solve the Pauli equation first, which charged particles satisfy in a magnetic field under symmetric gauge or Landau gauge, then to present energy levels, wave functions, expectation value of electron motion area and energy degeneracy so as to calculate Hall resistivity. But there are some demerits in the literatures. For example, Laughlin's theory  can not easily unify the formulation of the integer and the fractional quantum Hall effects. In addition, in order to interpret the fractional quantum effect the concepts of fractionally charged quasiparticles and quasihole etc. need to be proposed.

In this letter, Guided by concepts and theories of quantum mechanics and with the solution of the Pauli equation in a magnetic field under the symmetric gauge, wave functions and energy levels of single electrons, magnitude a of electron motion's spatial scope and the expectation value of electron motion area are first reviewed and presented. Then, after the quotation of non-interacting dilute gas system, the product of single electron's wave functions is used to construct wave functions of the electron gas system in a magnetic field, so as to define move area operator of the system and to calculate the expectation value of the system's motion area and electron degeneracy of the system. The letter is organized as follows. In next section  a brief review of wave functions and energy levels of single electrons in a magnetic field  is presented, in section 3, there are some  detailed discussions of Laughlin wave function in interaction electron gas system, in section 4, the wave functions and the energy levels of our model is presented explicitly, from which the electron Hall area density of the system is discussed. Finally, some remarks and conclusions are given in the last section.

\section{ A brief review of  the wave functions and energy levels of single electrons in a magnetic field}
Let's begin with a review of the wave functions and energy levels of single electrons in a magnetic field. From our previous  works\cite{WLL}-\cite{LWL}, we know that, after solving the Pauli equation, the normalized wave functions and the energy levels are given as follows.
\begin{equation}\label{probabilityarea}
\begin{array}{ll}
 \psi _{n_{\rho }m\, \lambda }(\rho ,\varphi \, ,\, z,\, s\, ,\, t\, )=(\frac{(n_{\rho }+|m|)!}{\pi a^2n_{\rho }!}){}^{1/2}e^{-\frac{\xi ^2}{2}}e^{i\, m\, \varphi }\\
\times e^{-i\, p_z\, z/\hbar }e^{-i\, E\, t/\hbar }\chi _{\lambda }(s)\underset{k=0}{\overset{n_{\rho }}{\, \sum }}\frac{(-1)^kC_{n_{\rho }}^k}{(k+|m|)!}\xi \, ^{2k+\, |m|},
\end{array}
\end{equation}
\begin{equation}\label{energyformula}
 E_{n_{\rho }m\, \lambda }=\frac{p_z{}^2}{2\mu }+(n_{\rho }+\frac{|m|+m+\lambda +1}{2})\frac{e\hbar B}{\mu c},
\end{equation}
in which $n_{\rho }=0,1,2,\text{...}\, ;\, m=0,\pm 1,\pm 2,\text{...}\, ;\, \lambda =-1,1,a=(2\hbar c/(eB))^{1/2},C_{n_{\rho }}^k=\text{Binomial}\left[n_{\rho },k\right]$ is combinatorial coefficient.

Straight forward calculation leads to the expectation value of electron motion area which reads
\begin{equation}\label{probabilityarea}
\begin{array}{ll}
  \langle S\rangle _{n_{\rho }m}=\pi \int _0^{2\pi }d\varphi \int _0^{\infty }\rho ^3|\phi (\rho ,\varphi ,z,s)|^2d\rho\\
    \\
 =\pi a^2\frac{(n_{\rho }+|m|)!}{n_{\rho }!}\sum _{k=0}^{n_{\rho }} \sum _{j=0}^{n_{\rho }} \frac{(-1)^{k+j}C_{n_{\rho }}^kC_{n_{\rho }}^j(|m|+j+k+1)!}{(|m|+k)!(|m|+j)!}\\
    \\
 =\pi (2n_{\rho }+|m|+1)a^2 .
\end{array}
\end{equation}
Obviously, the area is related to quantum number $n_{\rho }$ and $m$. In the physical condition when the quantum Hall effect appears, electrons are polarized by the magnetic field, then the direction of the electron magnetic moment and external magnetic field is the same (i.e. $\lambda =-1$). Only  $m\leq 0$ develops the lowest energy, and energy levels in (2) have nothing to do with angular momentum quantum number $m$ . In this case, the number of states is  $N_B=|m|+1$ whose quantum numbers are $n_{\rho }$ and $m$, so the degeneracy  per unit area (also named electron gas Hall surface density) is
\begin{equation}\label{probabilityarea}
\begin{array}{ll}
    n_B=\frac{N_B}{\langle S\rangle _{n_{\rho }m}}=\frac{|m|+1}{\pi (2n_{\rho }+|m|+1)a^2}=\, \nu \frac{eB}{hc}\\
  \nu =\frac{|m|+1}{2n_{\rho }+|m|+1},
    \end{array}
\end{equation}
Here, $\nu$ is our new defined quantum Hall number, which is named filling factor in the literature.  When
$n_\rho=0$, gives $\nu =1$, in this case,  the angular quantum number $m$ can still take one of $|m|+1$ values: $0, -1,-2 \cdot\cdot\cdot-|m|$, which gives the explanation the integer quantum Hall effects. When $n_\rho \neq 0$, the electrons of the electron gas system  stay in excited sates, different $n_\rho$ and different $|m|$ corresponding to different fractional quantum Hall number, all known fractional quantum Hall effects are included in this formulation, and what is more, this formulation can predicts some more fractional quantum Hall effects to be checked by experiment.

\section{ Laughlin wave function in interacting electron gas system}
Now, let's begin to discuss Laughlin wave function. In order to interpret quantum Hall effect, under the symmetric gauge,  Laughlin solved the Pauli equation for a single electron moving in a magnetic field by ladder-operator method and gave energy levels and wave functions of it as follows.

\begin{equation}\label{energylevel}
\begin{array}{ll}
    \psi _{n,m}(z,z^*)=\text{const}\cdot(b^{\dagger })^m(a^{\dagger })^n\psi _{0,0}(z,z^*)\\
   = \text{const}\cdot(z-\frac{\partial }{\partial z^*})^m(z^*-\frac{\partial }{\partial z})^ne^{-z^*z},
\end{array}
\end{equation}
\begin{equation}\label{}
E_n=(n+1/2)\text{ }\hbar \text{ }\omega  _c+\frac{p_z^2}{2\mu },\text{   }\omega  _c=eB/\mu c ,
\end{equation}
in which $z=\frac{x+iy}{2a},z^*=\frac{x-iy}{2a},\alpha=(\hbar c/eB)^{1/2}.$

Then, on the ground state Laughlin calculated the expectation value of electron motion area
\begin{equation}\label{energylevel}
\langle S\rangle _{0,m}=\pi (m+1)2\text{  }\alpha  ^2=(m+1)hc/eB .
\end{equation}

He estimated ground state degeneracy $m$ and gave ground state electron surface density
\begin{equation}\label{}
n_B=\frac{m}{\langle S\rangle _{0,m}}=\frac{m}{m+1}\frac{eB}{\hbar c}\approx \frac{eB}{\hbar c} .
\end{equation}

Because Laughlin did not get correct expectations formula of electron movement area
\begin{center}
$\langle S\rangle _{n,m}=\pi \left\langle \psi _{n,m}\left|\rho ^2\right|\psi _{n,m}\right\rangle$,\\
\end{center}
what he could do was to explain the integer quantum Hall effect with the theory.

On the other hand, in order to explain the fractional quantum Hall effect Robert B.Laughlin proposed interaction electron gas Hamiltonian operator which reads
\begin{equation}\label{}
H=\sum _j \frac{1}{2\mu }[\frac{\hbar }{i}\nabla _j+\frac{e}{c}\overset{\rightarrow }{A}){}^2+V(z_j)]+\sum _{j<k} \frac{e^2}{\left|z_j-z_k\right|},
\end{equation}
in which $V$ is the potential energy produced by charge surface density on uniform
 positive ion surface. Laughlin wrote the following wave function\cite{laughlin1}
\begin{equation}
\psi _{m\, }(z_1,z_2\, ,\, \cdots ,z_N\, )=\prod _{j<k}^N (z_j-z_k){}^m\, exp(-\frac{1}{4}\sum _{i=1}^{N} \left|z_i|^2\right.),
\end{equation}
in which $m$ is an odd-integral number meaning $j$th electron's angular momentum corresponding to $k$th electron is $m$. Thus, total angular momentum of N electron system is $N(N-1)m$/2. Here, $m$ is the filling factor used by electrons to occupy energy levels. Laughlin demonstrated $m=1/\nu$ . With this, some concepts, such as fractionally charged quasi-particle and quasi-holes need to be introduced to interpret the fractional quantum Hall effect.

\section{The wave functions of non interacting electron gas system}

This section mainly deals with the wave functions of non interaction electron gas system. It is theoretically successful to use single electron model to deal with the problems about electron gas system in a magnetic field because it can not only reasonably interpret both the integer and the fractional quantum Hall effects, but also predicate some undiscovered fractional quantum Hall effects. In practice, single electron model is successful because in electron gas system the electron's moving scale is, at least, ten thousands times as large as  Hydrogen atom's scale. So the interactions among electrons are too weak to be noticed. Supposing electrons move in z direction and a kinetic energy of z direction can be ignored, then we have, in magnetic field, the total Hamilton operator of electron gas system as follows,
\begin{equation}\label{}
H=\frac{1}{2\mu }\sum _{j=1}^N (\frac{\hbar }{i}\nabla _j+\frac{e}{c}\overset{\rightarrow }{A}){}^2 .
\end{equation}

The wave function $\psi _{N_{\rho }M\lambda }(\left\{\xi _k\right\},\left\{\varphi _k\right\},\left\{s_k\right\})$  must be the product of wave functions of N electrons which construct the electron gas system
\begin{equation}\label{energylevel}
\begin{array}{ll}
    \phi _{N_{\rho }M\lambda }(\xi _1,\text{...}\xi _N;\varphi _1,\text{...}\varphi _N;s_1,\text{...}s_N)\\
   =\phi _{N_{\rho }M\lambda }(\left\{\xi _k\right\},\left\{\varphi _k\right\},\left\{s_k\right\})\\
   =\prod _{k=1}^N \phi _{n_km_k\lambda _k}(\xi _k,\varphi _k,s_k),
\end{array}
\end{equation}

\begin{equation}\label{energylevel}
\begin{array}{ll}
    E_{N_{\rho }M\lambda }=\sum _{k=1}^N (n_k+\frac{\left|m_k\right|+m_k+\lambda _k+1}{2})\frac{e\hbar B}{\mu c}\\
    =(N_{\rho }+\frac{|M|+M+\lambda +N}{2})\frac{e\hbar B}{\mu c}.
\end{array}
\end{equation}
The normalized  wave functions of $k$'th $(=1\sim N)$ electron are
\begin{equation}\label{energylevel}
\begin{array}{ll}
   \phi _{n_km_k\lambda _k}(\xi _k,\varphi _k,s_k)=\frac{\sqrt{(n_k+m_k)!}}{a\sqrt{\pi n_k!}}\chi _{\lambda _k}(s_k)\\
    \\
   \times e^{i\, m_k\, \varphi _k}e^{\left.-\xi _k^2\right/2}\xi _k^{\left|m_k\right|}\sum _{j=0}^{n_k} \frac{(-1)^j C_{n_k}^j}{(j+\left|m_k\right|)!}\, \xi _k^{2j},
\end{array}
\end{equation}

\begin{equation}\label{energylevel}
N_{\rho }=\sum _{k=1}^N n_k,\text{   }M=\sum _{k=1}^N m_k,\text{  }\lambda =\sum _{k=1}^N \lambda _k\text{  }.
\end{equation}
In equation's (12) to (15), $k$'th electron's radial quantum number $n_k=0,1,2,\text{...}$, angular quantum number $m_k=0,\pm 1,\pm 2,\text{...}$, spin quantum number $\lambda _k=-1,1$ . In terms of permutation symmetry, the total wave function of the system needs also to make anti-symmetric summation and the function will be quite complex,  so the summation is not considered in this letter.

For example, the wave function for N electron system in ground state ( with spin quantum number $\lambda=-1$, radial quantum number $n_{\rho }=0$) reads
\begin{equation}\label{energylevel}
\begin{array}{ll}
   \psi _{0M\lambda }(\xi _1,\text{...}\xi _N;\varphi _1,\text{...}\varphi _N;s_1,\text{...}s_N)\\
    =\psi _{0M\lambda }(\left\{\xi _k\right\},\left\{\varphi _k\right\},\left\{s_k\right\})=\prod _{k=1}^N \phi _{0m_k\lambda _k}(\xi _k,\varphi _k,s_k)\\
  =\prod _{k=1}^N \left(a \pi \left|m_k\right|!\right){}^{-1/2}e^{\left.-\xi _k^2\right/2}\xi _k^{\left|m_k\right|}\chi _{\lambda _k}(s_k)e^{i\, m_k\, \varphi _k}.
\end{array}
\end{equation}

Now, let's define the total operator of N electron gas' moving area as
\begin{equation}\label{}
    S=\pi \sum _{k=1}^N \rho _k^2 .
\end{equation}
Then the expectation value of the  total moving area is
\begin{equation}\label{}
 \langle S\rangle _{N_{\rho }M}=(2N_{\rho }+|M|+N)\pi a^2 .
\end{equation}

When all $\text{\textit{$\left\{m_k\right\}$}}\leq 0$, for a certain energy level $E_{N_{\rho }M\lambda }=(N_{\rho }+\frac{\lambda +1}{2})\frac{e\hbar B}{\mu c}$, its degeneracy can be calculated from the sum of single electron's degeneracy and it is read $|M|+N$ .
 Thus, in N electron system,  the statistical average of electron surface density is
\begin{equation}\label{energylevel}
\begin{array}{ll}
   n_B=\frac{|M|+N}{(2N_{\rho }+|M|+N)\pi a^2}\\
  =\frac{\sum _{k=1}^N \left|m_k\right|+N}{2\sum _{k=1}^N n_k+\sum _{k=1}^N \left|m_k\right|+N}\frac{eB}{hc},~m_k=0,-1,-2,\text{...}\\
  \end{array}
\end{equation}
Only when all electrons occupy the same quantum number, that is $m_1=m_2=\text{...}m_N=m,\text{   }n_1=\text{...}n_N=n_{\rho }$ , can the following simple result (ever given by [1-2]) be obtained,
\begin{equation}\label{}
    n_B=\frac{|M|+N}{(2N_{\rho }+|M|+N)\pi a^2}=\frac{|m|+1}{2n_{\rho }+|m|+1}\frac{eB}{hc} ,
\end{equation}
which is same as equation(4).  Generally, with given energy level, all electrons do not occupy the same quantum number state. Equation (19) must be used to calculate the electron surface density for specified energy level in N electron system, but there should be a higher degeneracy and higher surface density than that given by (20). When all electron magnetic moments reverse the magnetic field (corresponds to $\left\{\lambda _{\text{\textit{$k$}}}\right\}=1$), for the energy levels given by both radial and angular quantum numbers, electron gas system surface density is
\begin{equation}\label{}
\begin{array}{ll}
  n_B=\frac{N_\rho+|M|+N}{(2N_{\rho }+|M|+N)\pi  a^2}\\
   \\
=\frac{\sum _{k=1}^N n_k+\sum _{k=1}^N m_k+N}{2\sum _{k=1}^N n_k+\sum _{k=1}^N \left|m_k\right|+N}\frac{eB}{hc},~ m_k=1,2,3,\text{...}
  \end{array}
\end{equation}

Moreover, for more general case, in electron gas system in a magnetic field, if some electrons' magnetic moments are in the same direction as the magnetic field(mark these electrons by $\uparrow$), others electrons'  magnetic moment's direction reverse to the direction of the magnetic field(mark these electrons by $\downarrow$). In this case,   the surface density of specified energy level will be
\begin{equation}\label{ }
   n_B=n_B^{\uparrow }+n_B^{\downarrow },\text{$\, $ } N=N^{\uparrow }+N^{\downarrow } ,
\end{equation}
\begin{equation}\label{}
\begin{array}{ll}
   n_B^{\uparrow }=\frac{\sum _{k=1}^{N^{\uparrow }} \left|m_k^{\uparrow }\right|+N^{\uparrow }}{2\sum _{k=1}^{N^{\uparrow }} n_k+\sum _{k=1}^{N^{\uparrow }} \left|m_k^{\uparrow }\right|+N^{\uparrow }}\frac{eB}{hc}, \\
   m_k^{\uparrow }=0,-1,-2,\text{...};~n_k=0,1,2,\text{...}\\
  \end{array}
\end{equation}
\begin{equation}\label{}
\begin{array}{ll}
   n_B^{\downarrow }=\frac{\sum _{k=1}^{N^{\downarrow }} n_k+\sum _{k=1}^{N^{\downarrow }} m_k^{\downarrow }+N^{\downarrow }}{2\sum _{k=1}^{N^{\downarrow }} n_k+\sum _{k=1}^{N^{\downarrow }} m_k^{\downarrow }+N^{\downarrow }}\frac{eB}{hc}\text{  }, \\
  m_k^{\downarrow }=1,2,\text{...};~n_k=0,1,2\text{...}\\
  \end{array}
\end{equation}

For specified energy level the Hall surface density in N electron gas system can be bigger(compared with single electron).  Equation (20) is only a special case.

\section{Some remarks and conclusions}

 Firstly we would like to provide similarities and differences between Laughlin wave function and wave functions in non-interaction electron gas system.  After a comparison of wave functions in non-interaction electron gas system of equation's (12) to (15) (we call it "our wave function" in the following ) and Laughlin wave function of (10), some remarks are in order.
(i) Our wave functions contain only the positive integer power of all electrons' dimensionless radial coordinate. While, electron wavelet functions contain both decaying exponential function and high order polynomial of electron coordinates. Our wave functions can develop the formula of Hall resistance which can be used to interpret both the integer and the fractional Hall effects without any suppose concepts.
(ii) Laughlin wave function also contains decaying exponential function of every electron coordinate. At the same time, it contains electrons' all possible odd-integer power's product. However, it can not develop the formula of Hall resistivity. When the wave function is really used to interpret integer and fractional quantum Hall effect, it has to ask the help from fractionally charged quasiparticle. In fact, it is not appropriate to declare the presence of fractionally charged quasiparticles in the system,  for there is no enough evidence.
(iii) Laughlin wave function is not the rigorous solution of Pauli equation of N interaction electron system, but a guess resulted from some hypothesis. In it $m$ is too artificial.
(iv) Our wave function, which is less artificial than Laughlin wave function, is the rigorous solution of Pauli equation of N non-interaction electron system under symmetric gauge, whose concise principles are easy to understand and accept so as to further the study and understanding of quantum mechanics.
(v) The formula of Hall surface density developed from the our wave function can provide satisfied interpretation for physical mechanism of quantum Hall effect. Hall surface density is closely related to quantum number $n $ and $m$. Namely, the different energy level corresponds to different surface density.   The more the radial quantum numbers are, the higher the energy is. Then the lower the surface density will be, the fewer electrons will be filled. Moreover, the surface density is in proportion to the external magnetic field in electron gas space. The stronger the field is , the greater $n$  is, then more electrons can be filled in specified level. Electrons in N-type semiconductor have a definite density. When the magnetic field is small, every electron level's surface density is also small. Then, fewer electrons will be filled. In fact, a number of electrons fill highly excited state so that more energies are produced with the interaction of electrons and crystal lattices. Thus, electron move's resistance increases. The greater the resistance is, the less opporunity for  superflow and superconductivity will appear.

When the magnetic field weakens, electron's Hall surface density $n_B$ decreases. Some electrons will shift to a space corresponding to a bigger radius probability orbit or even to high level energy state. Superflow and superconductivity appears only when electrons fill all the low-energy state probability orbit. In other words, if the low-energy state probability orbit is not completely filled with electrons when electrons are in a high-state level, it is impossible for superflow and superconductivity to appear.

In conclusion, the wave functions in non-interaction electron gas system are presented in this paper, and electron gas system surface density for different cases are discussed. Secondly the Laughlin wave function is reviewed.  Similarities and differences between two wave functions show that Hall surface density and wave functions developed from non-interaction electron model can explain the integer and the fractional quantum hall effects concisely and naturally. Moreover, the new formulation predicts that there may exit more other fractional quantum Hall effects.  Further studies on physical mechanism of the quantum Hall effect, such as its relation to Fermi-Dirac Condensed state, will be provided elsewhere.

\begin{acknowledgments}
This work was supported by the National Natural Science Foundation of China (10447005, 10875035),
an open topic of the State Key Laboratory for Superlattices and Microstructures (CHJG200902), the
scientific research project in Shaanxi Province (2009K01-54) and  Natural Science Foundation of Zhejiang Province (Y6110470),

\end{acknowledgments}


\end{document}